# Spillovers and Effect Attenuation in Firearm Policy Research in the United States

Lee Kennedy-Shaffer[*,1] and Alan H. Kennedy[2]

[*] Correspondence to: Lee.Kennedy-shaffer@yale.edu

[1] Department of Biostatistics, Yale School of Public Health, New Haven, CT, USA

[2] Program in Public Policy, The College of William & Mary, Williamsburg, VA, USA

**Abstract**

In the United States, firearm-related deaths and injuries are a major public health issue. Because of limited federal action, state policies are particularly important, and their evaluation informs the actions of other policymakers. The movement of firearms across state and local borders, however, can undermine the effectiveness of these policies and have statistical consequences for their empirical evaluation. This movement causes spillover and bypass effects of policies, wherein interventions affect nearby control states and the lack of intervention in nearby states reduces the effectiveness in the intervention states. While some causal inference methods exist to account for spillover effects and reduce bias, these do not necessarily align well with the data available for firearm research or with the most policy-relevant estimands. Integrated data infrastructure and new methods are necessary for a better understanding of the effects these policies would have if widely adopted. In the meantime, appropriately understanding and interpreting effect estimates from quasi-experimental analyses is crucial for ensuring that effective policies are not dismissed due to these statistical challenges.

Firearm-related deaths are a major public health challenge in the United States. Given the limited prospects for federal policies to address the issue, much of the policy development has fallen to the states. Evaluating state-level policies thus takes on elevated importance, creating an evidence base used by that state and others. In recent articles, Crifasi et al. discussed the current state of firearm policy research, focusing largely (though not exclusively) on state-level policies, and Schleimer et al. reviewed confounder selection practices in such research.[1,2] The authors compellingly and informatively review the current evidence on several policies and common methods. In addition, Crifasi et al. comment on several methodological and practical limitations, while also detailing an approach for controlled interrupted time series analysis that avoids some of the methodological limitations of other common quasi-experimental methods like difference-in-differences and synthetic controls.[1]

These methods, however, all rely on a key assumption: no spillover effects between the intervention and control states. This goes beyond adjusting for confounders and is often unstated in confounder-adjusted studies.[2] This assumption, and its consequent effects on bias and on the estimand targeted by these studies, poses a substantial challenge for such analyses. We discuss this challenge, its scale, and its statistical consequences. We then describe the state of current methodological approaches and identify research and data infrastructure needed to address the issue. Finally, we conclude with recommendations on appropriate interpretation and contextualization of policy evaluations.

**THE MOVEMENT OF FIREARMS**

The core challenge arises because individuals in a U.S. state are affected not just by their own state's gun laws but those of nearby states as well.[3,4] These nearby states may either be geographically contiguous or along natural routes for the movement of people and weapons such as the so-called "iron pipeline" northbound on Interstate 95 along the East Coast.[5]

Many studies have sought to describe the movement of guns—specifically guns used in crimes as recorded by the Bureau of Alcohol, Tobacco, Firearms and Explosives (ATF)—between states and how the relative strength of state laws affects that flow.[3–12] More challenging is determining the effect of neighboring state policies on crime rates and, especially, public health outcomes such as firearm deaths.[13–16] This issue has been noted as a limitation, along with the heterogeneous effects of policies, in the public health literature on firearm policies.[17–19]

This movement can happen at a large scale, with a state-level mean of 33% of traced crime guns coming from out-of-state between 2010 and 2019 in one recent study,[4] and a majority of state-years seeing over 100 crime guns from at least one other state in another study.[11] This movement is affected by state-level laws[4,10–12] and affects gun-related crimes and outcomes.[4,7,8]

**STATISTICAL CONSEQUENCES OF SPILLOVER EFFECTS**

This movement of firearms creates the opportunity for spillover effects of state-level policies, especially policies focused on limiting access to firearms. These spillovers or indirect effects have several important statistical consequences for a variety of analyses. We focus on their consequences for quasi-experimental panel data methods but much of the discussion applies as well to other regression-based or matching-based methods.

Most causal inference methods assume no interference: each unit's (i.e., state's) outcome should be affected by their own intervention status but not that of any other unit. This can be violated either: (1) by interventions affecting nearby control states; or (2) by the lack of intervention in control states affecting nearby intervention states.

The former is perhaps less likely in practice. Polarization around firearm policy has led in recent decades to states with pre-existing stronger gun laws strengthening them while states with weaker gun laws further loosened them.[4] So a state implementing stricter policies would probably be among the stricter states already and not a major source of guns elsewhere. If, however, a state policy did affect the control states (e.g., by reducing available guns in the region or limiting pipelines), those states' outcomes would not be true controls. They would be more like the intervention states, causing bias toward the null and reduced power. This phenomenon has been noted, for example, in infectious disease interventions, where there can be a large indirect effect of the treated units on the controls.[20]

The more likely consequence is that nearby control states will affect the intervention states by allowing individuals to bypass the effects of the policy; this phenomenon (variously termed "bypass", "offset", "leakage", or "perverse spillover") has been observed, for example, in state and local taxes on cigarettes and sugar-sweetened beverages.[21–23] Depending on the causal estimand of interest, this either induces bias towards the null (similar to the previous case), or changes the interpretation of the estimand itself. Quasi-experimental designs generally target the average treatment effect *on the treated* (ATT), meaning the attenuation of the effect is in fact part of the target estimand. That is, the method estimates the effect of this policy change in the specific state in which it occurred, holding fixed the policies of other states.[20]

This feature of the estimand limits generalizability and external validity.[20,24] If the same policy intervention occurred in a different state, specifically one with different nearby states or different policies affecting firearm movement, the effect could be either larger or smaller, potentially contributing to the heterogeneity observed in reviews of evidence.[2,23,25] If this policy has a true effect and was adopted nationwide, its effect could be noticeably larger by eliminating the possibility of effect attenuation by nearby states.[13] Moreover, these spillover effects can induce time-varying treatment effects, as the movement of firearms (and even other states' policies) may change in response to the intervention.[4] These second-order time-varying effects can affect the validity of results; methods that can handle time-varying effects must be used, further complicating interpretation.[1,20,26]

**EXISTING METHODS AND APPROACHES**

In recent years, statisticians and social scientists have developed a variety of methods to handle spillovers, with applications to infectious disease control,[27–29] educational interventions,[30–33] and place-based policies on taxes, economic development, and road safety, among others.[21–23,25,34]

With a known network through which spillover can occur, frameworks have been laid out for causal estimation through an "exposure mapping" that quantifies the ability of each unit's intervention to affect the other units.[32] This may involve defining blocks or clusters that are treated as independent or a more complex network that is treated as one of a super-population of networks or a part of a hypothetical larger network.[27,28,35]

For place-based policy effects, where spillover likely occurs not in discrete blocks but via distance-related networks, methods are also available. In much of the literature, the focus has been on: (1) defining the extent of the spillover effect from treatment units onto control units through a treatment effect on "those close to treatment"[34] or a "neighboring control"[22] or, more generally, spillover effects[21,23,25]; and (2) using that information to remove bias from the estimation of the ATT due to improper controls.[21,22,34] Two recent articles have developed estimands that explicitly account for the bypass effects by estimating the offset treated effect on the treated unit with untreated neighbors[23] or by estimating the effect of the full set of treatment conditions compared to no unit being treated.[25]

These methods can provide some understanding of the scope of the offsetting or spillover effect. However, they generally rely on replication across different units with differing neighboring exposure status and some homogeneity assumptions of the treatment effect across units. In addition, they rely on measurement of the spillover effect in the control units. For public health outcomes of firearm policies, this may not be available as the outcomes of interest occur in the treated states despite being caused by movement from untreated states. Finally, because more control units are observed than treated units for novel policies, most methods focus on comparisons to the full-control setting, rather than providing any means for estimating the effect if all units are treated. So the extent of underestimation of the policy effect remains unidentified.

Federal regulations provide a means to assess the full treatment effect, as in the case of ATF regulation of so-called "ghost guns" that was recently upheld by the Supreme Court. This federal policy sought to solve the issue of diverging state-level policies and the movement of firearms without serial numbers from one location to another.[36] These policies, however, face their own statistical challenges for analysis, as they require explicit modeling of counterfactual secular trends.

**AVENUES FOR PROGRESS: DATA AND METHODOLOGY**

To better understand the scope and scale of this challenge, accurate measures of the movement of firearms and ways in which this movement undermines state-level policies are necessary. Existing research often uses crime gun statistics from ATF, relying on Freedom of Information Act requests that can delay and limit data availability.[4] As discussed by Crifasi et al.,[1] these data are difficult to integrate with other sources. Because ATF only traces guns used in crimes, and many are unsuccessfully traced, missing data problems arise here as well. In recent years, gun kits without serial numbers ("ghost guns") exacerbated this problem, although that may have declined with recent ATF regulations.[36] An ongoing, maintained database with more current information would be an important start to addressing this challenge, but additional data and tracing are needed to fully account for the role of firearm movement. Spillover measures are also better defined on more granular levels where cross-border effects can be more clearly seen.

Additional information and research expanding on these movement patterns can inform the networks that are necessary to define exposure mappings of policies.[32] This can be useful for identifying units subject to spillover and bypass effects, which can explain potential biases. The movement of guns is a proxy, and potentially mediator, of the spillover and bypass effects of policies on public health outcomes, so databases that combine those two sources of information would be valuable for building models of that relationship that can be incorporated into policy evaluation.

Adapting recent methods that identify the total effect of a treatment regime compared to no treatment of any unit can improve estimation of these effects. To generalize estimates to the effect of national implementation of a policy (or, e.g., regional implementation of coordinated policies), however, requires further development of methods and, likely, parametric models of direct and indirect effects. This requires building more complete models of movement patterns and how they are affected by policy changes, as well as how they affect public health outcomes.

**CONCLUSION**

The statistical challenges laid out here limit the ability of states to serve as "laboratories" for social policies, as Justice Louis Brandeis suggested in 1932.[4] Policies that truly are effective may have underestimated effects or underpowered hypothesis tests and thus be inaccurately labeled ineffective. A useful policy implemented in only one or a handful of states may not demonstrate its full effectiveness and thus be set aside by policymakers or advocates demanding statistical evidence.[37]

The concurrent development of data infrastructure and methods is crucial to understanding the effects of scaling up policies. Much work is required to develop this framework generally and to apply it specifically to firearm policy. In the meantime, however, estimates from existing methods must be interpreted properly to ensure statistical limitations do not inhibit the adoption of effective policies. Interpretations from approaches that do not account for spillover should clearly lay out the risks of bias from spillover effects and describe the specific estimand targeted, with reference to existing understanding of the movement of guns. Discussion should include how the treatment effect might change if the policy is implemented at a different scale or in a place with different avenues for bypass (i.e., surrounding state policies). Understanding and communicating these challenges and the potential for larger-scale effects can provide more meaningful long-term policy understanding and more effective use of statistical evidence.